# Designing AI for Real Users: Accessibility Gaps in Retail AI Front-End

Neha Puri and Tim Dixon
Intertek Group Plc

**ABSTRACT**

As AI becomes embedded in customer-facing systems, ethical scrutiny has largely focused on models, data, and governance. Far less attention has been paid to how AI is experienced through user-facing design [11]. This commentary argues that many AI front-ends implicitly assume an "ideal user body and mind," and that this becomes visible—and ethically consequential—when examined through the experiences of differently abled users. We explore this through retail AI front-ends for customer engagement – virtual assistants, virtual try-on systems, and hyper-personalised recommendations. Despite intuitive and inclusive framing, these systems embed interaction assumptions that marginalise users with vision, hearing, motor, cognitive, speech, and sensory differences, as well as age-related variation in digital literacy and interaction norms. Drawing on practice-led insights, we argue that these failures persist not primarily due to technical limits, but due to the commercial, organisational, and procurement contexts in which AI front-ends are designed and deployed, where accessibility is rarely contractual [15, 14]. We propose front-end assurance as a practical complement to AI governance, aligning claims of intelligence and multimodality with the diversity of real users.

## 1 Introduction

AI front-end design increasingly mediates everyday human–AI interaction, yet ethics discussions focus primarily on models, data, and governance rather than how people experience AI systems at the interface [11]. This commentary argues that many contemporary AI front-ends implicitly assume an "ideal user body and mind," and that this assumption becomes visible—and ethically consequential—when examined through the experiences of differently abled users. Differently abled users do not represent edge cases; they reveal where systems have been designed around a narrow and idealised model of human ability. The World Health Organization estimates that over one billion people globally live with some form of disability [16].

Our contribution is a practice-led analysis that links front-end exclusion not to technical limits, but to organisational, procurement, and regulatory structures. We focus on retail AI front-ends used for customer engagement—conversational assistants, virtual try-on, and hyper-personalised recommendations—and show how interaction assumptions marginalise users with vision, hearing, motor, cognitive, speech, and sensory differences, as well as age-related variation in interaction norms. We then examine why these failures persist and propose front-end assurance as a missing governance layer for ethical AI.

## 2 The "Ideal User" as a Default Design Assumption

### 2.1 Real-world design drivers

Front-end interaction design is shaped by market optimisation, vendor defaults, and inherited visual paradigms. Success metrics such as engagement and conversion tend to prioritise sighted and neuro-typical interaction patterns [15]. Adaptive AI components, often procured as third-party modules, are integrated "as-is," with accessibility features optional rather than enforced [14]. As accessibility is rarely specified in procurement or tested in deployment, inclusive interaction is treated as a late-stage retrofit rather than a design requirement. These assumptions also extend to age: interfaces are frequently described as "intuitive" based on familiarity within a narrow age band, implicitly assuming shared knowledge of interaction patterns, symbols, and gestures that may not hold for younger or older users.

## 2.2 Regulatory drivers

The European Union is currently ahead of other regions in mandating digital accessibility. The European Accessibility Act (EAA) establishes a harmonised, legally binding baseline for accessible digital products and services across the EU from 28 June 2025, including e-commerce and customer-facing platforms [5]. However, these obligations apply to digital services in general rather than to AI systems specifically.

In contrast, the EU Artificial Intelligence Act (AI Act) adopts a risk-based approach and does not directly apply to most customer-facing retail AI front-ends. Its obligations phase in over time and primarily target high-risk AI systems [6].

## 2.3 Implications

Together, these dynamics mean that while brands are under immediate accessibility obligations for their digital interfaces under the EAA, there is currently limited AI-specific regulatory pressure to ensure accessible AI front-ends. In this gap, organisational incentives and vendor defaults embed a narrow "ideal user" model into AI interfaces—even where inclusive interaction is technically possible.

# 3 Illustrative Ethical Fault Lines in Retail AI Front-Ends

The ideal user assumption materialises in three widely deployed retail AI front-ends; across them, assumptions about sight and speed are compounded by assumptions about age-specific familiarity with interface conventions.

## 3.1 Virtual Try-On: Exclusion by Design

Virtual try-on tools based on computer vision and generative models are widely deployed by apparel and eyewear brands to simulate fit and appearance and reduce returns [1]. These systems are often presented as improving accessibility by enabling remote shopping for users who face barriers to visiting physical stores.

However, current implementations remain predominantly visual, relying on photorealistic overlays and body or face mapping. Blind and low-vision users—and some older users unfamiliar with gesture- or image-driven interfaces—cannot access, query, or verify what the system is "showing," revealing a default assumption of sight that is neither tested nor audited at the interface. This illustrates a broader ethical tension: an intervention that improves access for some users can simultaneously create new forms of exclusion for others.

This limitation is not inherent to AI. Voice-first assistants can provide structured descriptions of products (e.g., fit, colour, fabric behaviour) and support non-visual shopping journeys [2]. Integrating voice or text-based descriptions of try-on outputs, alongside auditable semantic metadata describing what the model inferred and why, would allow these systems to extend accessibility benefits more equitably.

## 3.2 Hyper-Personalised Ranking: Invisible Coercion

Retail platforms increasingly use machine learning to dynamically reorder products and highlight "recommended," "popular," or "limited time" items [10]. These signals are conveyed through visual hierarchies, badges, and colour cues. Age further complicates this dynamic: such cues are often treated as self-explanatory, yet their meaning may be opaque to older users unfamiliar with platform-specific conventions, or to younger users encountering them outside their expected context. In many deployed systems, these cues are not programmatically exposed to assistive technologies. Accessibility audits of AI chat and recommendation widgets report missing semantic markup, poor focus management, and unannounced content updates that obscure the reasons behind ranking [14, 12]. What appears as a subtle nudge for some users becomes a command by default for others who cannot perceive the persuasion cues.

These effects can be mitigated through semantic labelling of ranking cues, accessible explanations of promotion logic, and evaluation with assistive technologies and diverse users.

## 3.3 Conversational AI for Support & Redress

AI chatbots and voice assistants now mediate customer service, returns, and complaints for many retailers [4]. Common accessibility barriers include unlabeled controls, keyboard traps, focus jumping, and screen readers failing to announce new messages [14]. While conversational interfaces are often positioned as more accessible, they can still assume age-specific norms around phrasing, pacing, and turn-taking that are not equally intuitive across generations.

When access to refunds, complaints, or escalation depends on navigating an inaccessible interface, routine support becomes a rights failure. These barriers can be countered using established inclusive design patterns: robust focus management, reliable announcements of new content, fully keyboard-navigable controls, and a clearly labelled human escalation path.

Across these fault lines, differently abled users expose where AI front-ends assume sight, rapid cognition, and visual persuasion. The countermeasures above reframe the interface as a site of ethical accountability rather than a neutral delivery layer.

## 4   Why These Failures Persist

These failures persist due to a combination of structural and governance factors:
- Performance optimisation metrics: AI systems are trained and evaluated on majority-user behaviour (e.g., click times, dwell times, conversion), embedding narrow assumptions about age, ability, and interaction speed.
- Inherited design systems: Retail front-ends often reuse interaction components from mainstream consumer platforms (badges, colour cues, infinite scroll), without inclusive reevaluation. For example, Microsoft Windows Copilot has exhibited a persistent issue where screen readers announce only part of a response, with spoken output diverging from the visible chat history and leaving blind users with incomplete or conflicting information [13].
- Procurement norms: Off-the-shelf AI modules are widely adopted, yet accessibility is rarely a contractual requirement or acceptance criterion.
- Regulatory ambiguity: Laws like the EAA apply broadly to accessible products and services, but guidance for adaptive and generative systems is still emerging, creating uncertainty for teams implementing AI features.

This misalignment between technical capability, organisational practice, and regulatory interpretation allows exclusionary design assumptions to persist.

## 5   Front-End Assurance: Towards Ethical Interfaces

Inclusive AI front-ends are technically feasible when accessibility is treated as a primary design goal rather than a retrofit. Applications such as Be My Eyes and InnoSearch.ai demonstrate how conversational interaction can support blind and low-vision users in interpreting images, asking follow-up questions, searching across retailers, and placing orders through accessible web or phone-based AI interfaces [3, 7].

At the same time, these intermediary solutions expose an ethical tension: by compensating for inaccessible retail platforms, they risk becoming a substitute for—rather than a catalyst for—accessible design by primary brands.

If the interface is where ethical consequences emerge, it must also be a site of accountability. We propose front-end assurance as a practical extension of AI governance, encompassing:
• Access testing: Can differently abled users perceive explanations, consent controls, and interaction alternatives?;
• Transparency checks: Are uncertainty and persuasion cues semantically exposed?;
• Agency audit: Can users override defaults or request human intervention?; and
• Consistency evaluation: Do disclosures and options vary with personalisation?

As an illustration, access testing could draw on established AI fairness metrics frameworks (e.g., [8, 9]), adapting group-based disparity measures from model outputs to interaction outcomes. (e.g. as task completion parity, error rate balance across ability groups). However, this adaptation requires careful attention to group taxonomy—disability categories are more fluid and intersectional than typical protected attributes such as gender or origin—and to data collection methodology, as interaction-level metrics require representative user testing rather than computational evaluation on held-out datasets. Developing standardised protocols for accessibility fairness measurement represents a valuable direction for future work.

Overall, treating front-end behaviour as an assurance surface—testable, measurable, and enforceable—is important to ensure that claims of intelligence, personalisation, and multimodality hold for the diversity of real users.

## 6   Conclusion

AI front-ends increasingly determine how people encounter brands, make choices, and seek support. Differently abled users reveal the narrow assumptions embedded in these interfaces—assumptions about sight, hearing, movement, and cognition that shape who can meaningfully engage with systems presented as intelligent or multimodal. When interaction works only for a subset of users, adaptability and intelligence become performative rather than real.

These assumptions are often justified through claims of "intuitive" design, yet intuitiveness is frequently age-specific and culturally learned rather than universal.

Importantly, accessibility is not a niche concern. Design choices that improve accessibility—clear explanations, consistent interaction patterns, multimodal feedback, and user control—often improve usability, trust, and resilience for all users, including those in situational or temporary states of limitation. Given that a significant proportion of the global population experiences disability at some point in their lives, inclusive front-end design is inseparable from designing for real-world use at scale [16].

Ethical AI cannot be achieved without ethical interfaces. Shifting accessibility from a late-stage accommodation to a core design, procurement, and governance requirement is therefore foundational. Front-end assurance offers a practical pathway to make claims of intelligence, personalisation, and multimodality meaningful by aligning AI systems with the diversity of real users.

## ACKNOWLEDGMENTS

This paper was informed by a Vision Awareness Seminar conducted by Tim Dixon, Head of IT Architecture at Intertek Group plc in November 2025. The seminar introduced a user-lens grounded in lived experience of vision loss, highlighting interaction challenges and potential design responses relevant to AI-enabled systems. This perspective helped shape the analytical framing adopted in this work.